\documentstyle[preprint,aps,psfig]{revtex}

\def\lsim{\raise0.3ex\hbox{$\;<$\kern-0.75em\raise-1.1ex
\hbox{$\sim\;$}}}
\def\gsim{\raise0.3ex\hbox{$\;>$\kern-0.75em\raise-1.1ex
\hbox{$\sim\;$}}}
\begin{document}
\textheight = 23.3cm
\topmargin = -1.6cm

\baselineskip 6.3mm
\begin{flushright}
\vglue -1.0cm
TMUP-HEL-0019 \\ 
KIAS-P00067\\ 
hep-ph/0010240 \\ 
\end{flushright}
\begin{center}
\Large\bf
Inverted Hierarchy of Neutrino Masses Disfavored by Supernova 1987A
 \end{center}
\begin{center}
Hisakazu Minakata$^{1,2,3}$ 
\footnote[2]{E-mail: minakata@phys.metro-u.ac.jp}
and Hiroshi Nunokawa$^{3,4}$
\footnote[2]{E-mail: nunokawa@ifi.unicamp.br}\\
\vskip 0.2cm
{$^1$\it Department of Physics, Tokyo Metropolitan University \\
1-1 Minami-Osawa, Hachioji, Tokyo 192-0397, Japan\\
\vskip 0.2cm
$^2$Research Center for Cosmic Neutrinos, 
Institute for Cosmic Ray Research, \\ 
University of Tokyo, Kashiwa, Chiba 277-8582, Japan}
\vskip 0.2cm
{\it $^3$School of Physics, Korea Institute 
for Advanced Study
\\
207-43, Cheongryangri-dong, Dongdaemun-gu, 
Seoul 130-012, Korea} \\ 
\vskip 0.2cm
{\it $^4$Instituto de F\'{\i}sica Gleb Wataghin, 
Universidade Estadual de Campinas - UNICAMP\\
P.O. Box 6165, 13083-970 Campinas SP Brazil} \\ 
(October 2000) 
\end{center}
\vspace{-1.0cm}
\begin{abstract}
We discuss the flavor conversion of supernova 
neutrinos in the three-flavor mixing scheme 
of neutrinos. 
We point out that by neutrino observation 
from supernova one can discriminate the inverted 
hierarchy of neutrino masses from the normal one if 
$s_{13}^2 \gsim \text{a few} \times 10^{-4}$, 
irrespective of which oscillation solution to the solar 
neutrino problem is realized in nature. 
We perform an analysis of data of SN1987A and obtain a 
strong indication that the inverted mass hierarchy is 
disfavored unless 
$s_{13}^2 \lsim \text{a few} \times 10^{-4}$. 
\end{abstract}
\newpage
The phenomenon of neutrino flavor transformation which was discovered 
in atmospheric neutrino observation by Kamiokande and Superkamiokande 
experiments \cite {SKatm} implies new neutrino properties, tiny masses 
and unexpectedly large mixing angles. 
It constitutes, at present, the unique evidence for physics beyond 
the standard model of particle physics. 
The persistent discrepancy between the observed and the calculated 
flux of solar neutrinos \cite {solar} provide another evidence 
in favor of these new neutrino properties. It is worth to note that  
these robust evidences perfectly fit into the standard three-flavor 
mixing scheme of neutrinos. Recently it is reported that the third 
member, the $\tau$ neutrinos, are experimentally detected for the 
first time \cite{DONUT}. 

Yet, there still remain many unknowns in the properties of neutrinos.
Among other things we have no experimental clue on the question of 
neutrino mass pattern, i.e., the normal vs. the inverted mass 
hierarchies, either from any terrestrial experiments or solar 
and atmospheric neutrino observation. Here, we mean, by normal and 
inverted mass hierarchies, the mass pattern 
$m_3 \gg m_1 \sim m_2$ and $m_1 \sim m_2 \gg m_3$,
respectively. In our notation 
$\Delta m^2_{ij} \equiv m^2_j - m^2_i$, and $\Delta m^2_{12}$ and
$\Delta m^2_{13} \sim \Delta m^2_{23}$
are the mass squared differences that are related with the 
solar and the atmospheric neutrino oscillations, respectively. 

The issue of normal vs. inverted mass hierarchies is related 
to the deep question of under what discipline nature organizes 
the neutrino mass spectrum. If the inverted hierarchy is the case 
the discipline is clearly quite different from that used to organize 
the quark sector. Determination of which mass hierarchy is 
realized should imply powerful constraints on 
model building of neutrino masses and mixing. 

Experimentally the problem of normal vs. inverted mass hierarchies 
is one of the key issue regarding the question of whether one can 
measure the absolute neutrino masses. 
The only imaginable way of detecting neutrino masses 
of the order of 
$\sqrt{\Delta m^2_{atm}} \simeq (0.04-0.07)$ eV
is the neutrinoless double beta decay experiments, 
which is of course possible only for Majorana neutrinos.
If the normal mass hierarchy is the case, one would have to 
go down to an order of magnitude smaller value of 
$<m_{\nu}>$, the observable parameter in the experiments, 
to $\sim 0.001 $eV \cite{doublebeta,Klapdor00}.
                               
In this paper we point out that neutrino observation from 
a galactic supernova can judge whether the normal or the inverted 
mass hierarchies is realized in nature, unless the mixing angle 
$\theta_{13}$ is extremely small, 
$s^2_{13} < \text{a few} \times 10^{-4}$.
Moreover we perform an analysis of the data of SN1987A in 
Kamiokande and IMB detectors \cite {SN1987A} and obtain a strong 
indication that it disfavors the inverted mass hierarchy.

Of course, this conclusion must be checked against the 
direct determination of the sign of $\Delta m^2_{13}$ which 
will be done in future long-baseline accelerator experiments 
\cite {JHF,MINOS,OPERA}.
However, the result we describe in this paper appears to be 
the unique hint which is available before such experiments 
are actually done. 

We start by summarizing the common knowledges on neutrinos from 
supernova (SN) \cite {Suzuki}
and their properties inside neutrinosphere 
\cite {MWS,SNsimu,Janka}. 

\noindent
(1) Consideration of energetics of SN collapse indicates that 
almost all ($\sim 99 \%$) of the gravitational binding energy of 
neutron star is radiated away via neutrino emission. The total 
energy is estimated to be several $\times 10^{53}$ erg, and  
it is expected that the energy is equipartitioned into 
three flavors in a good approximation \cite {MWS,B87}.

\noindent
(2) It is discussed that the shape of the energy spectra of 
various flavor neutrinos can be described by a "pinched"
Fermi-Dirac distribution \cite {JH89}. 
The pinched form may be parametrized by 
introducing an effective "chemical potential". 

\noindent
(3) There is no physical distinction between $\nu_{\mu}$ and 
$\nu_{\tau}$ and their antiparticles in neutrinosphere. 
It is because $\nu_{\mu}$ and ${\bar{\nu}}_{\mu}$ 
are not energetic 
enough to produce muons by the charged current interactions, 
and the neutral current cross sections of $\nu$ 
and $\bar{\nu}$ 
are similar in magnitude. 
Therefore, we collectively denote them as "heavy neutrinos"
in this paper.\footnote
{The terminology implicitly assumes that the normal mass hierarchy 
is the case. Nevertheless, we will use it even when we discuss the 
inverted mass hierarchy.}

\noindent
(4) The location of neutrinosphere of heavy neutrinos, $\nu_{\mu}$ 
and $\nu_{\tau}$, is believed to be in deeper place than 
$\bar{\nu}_{e}$ and $\nu_{e}$ in SN. 
It is due to the fact that the heavy neutrinos have weaker 
interactions with surrounding matter; they interact with matter 
only via the weak neutral current, whereas, 
$\bar{\nu}_{e}$ 
and $\nu_{e}$ do have additional charged current interactions. 
Hence, the heavy neutrinos have to have deeper neutrinosphere 
because their trapping requires higher matter density compared 
to those required for 
$\bar{\nu}_{e}$ and $\nu_{e}$. 

This last feature is of crucial importance for our business.
It implies that the heavy neutrinos are more 
energetic when they are radiated off at 
neutrinosphere because the temperature is higher 
in denser region. It may be characterized by the 
temperature 
ratios of $\nu_e$ and $\bar{\nu}_e$ to $\nu_{heavy}$
\begin{equation}
\tau \equiv 
\frac{T_{\nu_h}}{T_{\bar{\nu}_e}} \simeq
\frac{T_{\bar{\nu}_h}}{T_{\bar{\nu}_e}} \simeq
1.4-2.0
\end{equation}
according to the simulation of supernova dynamics which is 
carried out in Ref. \cite{MWS,SNsimu,Janka}.
Despite possible slight difference it should be 
a reasonable approximation of ignoring temperature 
difference between $\bar{\nu}_h$ and $\nu_h$. 

We now turn to the the neutrino flavor conversion in supernova (SN), 
the core matter in our discussion in this paper. In fact, it 
has a number of characteristic features which makes SN unique 
among other astrophysical and terrestrial sources.

\noindent
(i) Because of extremely high matter density inside neutrinosphere 
all the neutrinos with cosmologically interesting mass range, 
$m_{\nu} \lsim 100$ eV, are affected by the MSW effect 
\cite {MSW}. 
(Earlier references on the MSW effect in supernova include 
Ref. \cite{SNMSW}.)
Consequently, the three neutrino and three antineutrino 
eigenstates have two level crossings, first at higher (H) density 
and the second at lower (L) density, inside SN as 
schematically indicated in Fig. 1.

\noindent
(ii) The key question in the neutrino flavor conversion in 
SN is whether the H level crossing is adiabatic or not. If it is 
adiabatic, then the physical properties of neutrino conversion 
is simply $\nu_{e}-\nu_{heavy}$ exchange in the normal mass 
hierarchy. It should be emphasized that this feature holds 
irrespective of the possible complexity of the solar neutrino 
conversion which governs the L resonance. These key features 
have been pointed out in our earlier paper, Ref. \cite {MN90}.

\noindent
(iii) The second important question is if the neutrino mass 
spectrum adopts the normal or inverted mass hierarchies. 
If the mass hierarchies is of normal (inverted) type, 
the H level crossing is in the neutrino (antineutrino) 
channel. 

For a recent comprehensive treatment of neutrino flavor 
conversion in SN in the framework of three-flavor mixing, 
see Ref. \cite {SD99}.

The last two remarks are crucial in our business. It will 
allow us to determine which mass hierarchy is realized 
by analyzing neutrino data from SN without knowing the 
parameters in the solar neutrino solution. Notice that 
this statement is valid not only for the MSW but also for 
the vacuum solar neutrino solutions.

Before going on, let us elaborate (ii) because it is of crucial 
importance in our argument. 
The three solid lines at the right most end of the level crossing 
diagram, Fig. 1a, represent three neutrino mass eigenstates in 
matter at the core of SN. The neutrino evolution in SN obeys 
the Schr\"edinger-like equation
\begin{equation}
i\frac{d}{dx} 
\left[
\begin{array}{c}
\nu_e \\ \nu_\mu \\ \nu_\tau
\end{array}
\right] 
=
\left\{U \left[
\begin{array}{ccc}
m_1^2/2E & 0 & 0 \\
0 & m_2^2 /2E & 0 \\
0 & 0 & m_3^2/2E 
\end{array}
\right] U^{+}
+
\left[
\begin{array}{ccc}
a(x) & 0 & 0 \\
0 & 0 & 0 \\
0 & 0 & 0
\end{array}
\right]\right\}
\left[
\begin{array}{c}
\nu_e \\ \nu_\mu \\ \nu_\tau
\end{array}
\right],
\label{evolution1}
\end{equation}
where $a(x) = \sqrt{2} G_F N_e(x)$ indicates the index of
refraction with $G_F$ and $N_e(x)$ being the Fermi constant and 
the electron number density, respectively. In (\ref{evolution1})
$U$ denotes the leptonic flavor mixing matrix, the 
Maki-Nakagawa-Sakata (MNS) matrix \cite{MNS}. 

Because of an extreme density near the neutrinosphere, 
$\gsim 10^{10}$ g/cm$^3$, 
the matter term diag[$a(x)$, 0, 0] $\equiv H_0$ must be dominant 
over the other term. One can then formulate perturbation theory 
in which one takes the matter term $H_0$ as unperturbed part and 
the other one as perturbation. It is a degenerate perturbation 
theory and one has to diagonalize the 2 $\times$ 
2 subspace to obtain the zeroth-order wave functions 
and the first-order correction to the energy eigenvalues. 
The resulting three matter-mass eigenstates 
at neutrinosphere in the case of normal mass 
hierarchy are as follows: 

\vskip 0.3 cm
\noindent
$\nu^{m}_{3}$ is almost pure $\nu_e$. 
$\nu^{m}_{1}$ and $\nu^{m}_{2}$ 
are certain superposition of heavy neutrinos, 
$\nu_{\mu}$ and  $\nu_{\tau}$, but 
with negligible $\nu_e$ component. 

\vskip 0.3 cm

The left end of Fig. 1a describes the three 
vacuum mass eigenstates 
$\nu_{1}$, $\nu_{2}$ and $\nu_{3}$. Therefore, if the H resonance 
is adiabatic, the $\nu_{3}$ state has the same physical properties, 
e.g., temperature, as $\nu_e$ inside neutrinosphere. 
By unitarity, the properties of $\nu_{1}$ and $\nu_{2}$ 
states in vacuum, that is when they get out from the star, 
must be determined equally accurately by physical properties of 
$\nu_{\mu}$ and $\nu_{\tau}$ inside neutrinosphere. 
Simce there is no physical distinction between $\nu_{\mu}$ 
and $\nu_{\tau}$ there, the properties of $\nu_{1}$ and 
$\nu_{2}$ are not affected by the nature of the L level crossing, 
adiabatic, moderately nonadiabatic, or nonadiabatic.  
We suspect that this feature remains true even when the two 
level crossings come close so that the independent two-resonance 
approximation may not be completely valid. 

This completes the argument to show that if the H resonance 
is adiabatic the net effect of the neutrino flavor conversion 
in SN is simply the $\nu_e$-$\nu_{heavy}$ exchange, 
irrespective of the nature of the L level crossing.\footnote
{One can of course confirm this result by putting $P_{H}=0$ 
in the relevant equations in Ref. \cite{SD99}, where $P_H$ 
denotes the nonadiabatic ``jump'' probability. 
}
This is nothing but the feature that is called as 
"$\nu_e$-$\nu_{\tau}$ exchange" in Ref. \cite{MN90}.

The adiabaticity of the H resonance is guaranteed if 
the following adiabaticity parameter $\gamma$ 
is much larger than unity at the resonance point:
\begin{eqnarray}
\gamma &&\equiv \frac{\Delta m^2}{2E}
\frac{\sin^2 2\theta}{\cos 2\theta}
\left|\frac{\text{d}\ln N_e}
{\text{d}r}\right|^{-1}_{res}\nonumber\\
&&= 
\left(\frac{\Delta m^2}{2E}\right)^{1-1/n}
\frac{\sin^2 2\theta}{(\cos 2\theta)^{1+1/n}}
\ \frac{r_\odot}{n}
\left[\frac{\sqrt{2}G_F\rho_0 Y_e}{m_p}\right]^{1/n}, 
\end{eqnarray}
Here, we assumed that the density profile of 
the relevant region of the star can be described 
as $\rho(r) = \rho_0(r/r_\odot)^{-n}$ 
to obtain the second line in the above equation, 
where $r_\odot = 6.96 \times 10^{10}$ 
cm denotes the solar radius. 
With the choice $n=3$ and 
$\rho_0 \simeq 0.1$ g/cc \cite{Nomoto}, 
we get, 
\begin{equation}
\gamma \simeq 0.63 \times 
\left[\frac{\sin^2 \theta_{13}}{10^{-4}}\right]\ 
\left[
\frac{\Delta m^2}{10^{-3} \text{eV}^2} 
\right]^{2/3}
\left[
\frac{E}{20\ \text{MeV}}
\right]^{-2/3},  
\end{equation}
for the small value of $\theta_{13}$. 
Since the conversion probability $P$ is 
approximately given by 
$P=\exp[-\frac{\pi}{2}\gamma]$, 
$s_{13}^2 \gsim \text{a few} \times 10^{-4}$
assures adiabaticity in a good accuracy.

Now we notice that the basic elements for the argument toward 
disfavoring inverted mass hierarchy is actually very simple.
Because of (iii), the resonance is in the antineutrino channel 
if the inverted mass hierarchy is the case
as illustrated in Fig. 1b. 
It means that, 
if the H resonance is adiabatic, all the 
$\bar{\nu}_e$'s at 
neutrinosphere are converted 
into heavy antineutrino states, 
and vice versa. 
It is also known that if H resonance is 
adiabatic, final $\bar{\nu}_e$ spectrum 
at the detector is not affected by the earth matter effect 
\cite {SD99}.\footnote
{The reason is as follows. Let us first
note that $\bar{\nu}_3$ state which carry the
original $\bar{\nu}_e$ spectrum oscillate
very little into $\bar{\nu}_e$ in the earth 
because $|\Delta m_{13}^2|/E$ 
is much larger than the earth matter potential 
and also because $\theta_{13}$ is small \cite {CHOOZ}.
Therefore, the oscillation in the earth 
takes place essentially only 
between $\bar{\nu}_1$ and $\bar{\nu}_2$,  
decoupling the $\bar{\nu}_3$ state. 
It would lead to regeneration of $\bar{\nu}_e$ but it 
would not give any significant effect for the 
$\bar{\nu}_e$ component at the detector because both 
$\bar{\nu}_1$ and $\bar{\nu}_2$ carry 
original energy spectrum of heavy flavors 
at the neutrinosphere.}

Since the $\bar{\nu}_e$-induced charged current 
reaction is dominant in water Cherenkov detector, one can severely 
constrain the scenario of inverted mass hierarchy by utilizing 
this feature of neutrino flavor transformation in SN.
When the next supernova event comes it can be used 
to make clear judgement on whether the inverted mass hierarchy is 
realized in nature, a completely independent information from 
those that will be obtained by the long-baseline neutrino 
oscillation experiments. 

While waiting for the next galactic SN, let us perform an analysis 
of the data of neutrinos from SN1987A to gain a hint to the 
problem of the mass pattern which we want to solve. 
In the following analysis, we assume that $s_{13}$ 
is large enough, $s_{13}^2 \gsim \text{a few} 
\times 10^{-4}$,  
to guarantee the adiabaticity of the H resonance. 

In fact, very similar analyses have been done by several 
authors \cite {SSB94,JNR96}. 
We may be able to characterize 
our work in comparison with theirs in 
the following way; 
We formulate the problem in a proper 
setting of the three-flavor mixing scheme 
of neutrinos, which is essential 
for the SN neutrinos. In due course, we 
will try to clarify how conclusions obtained in earlier 
works are to be interpreted, or to be conditioned in the 
three-flavor framework. 

We follow Jegerlehner, Neubig and Raffelt 
\cite {JNR96} who employed the method of maximum likelihood. 
We define the Likelihood function as follows \cite {JNR96}: 
\begin{equation}
{\cal L}=C\, \exp 
\left(-\int_0^\infty n(E)dE \right)
\prod_{i=1}^{N_{\rm obs}} n(E_i),
\end{equation}
where $N_{\rm obs}$ is the total number 
of experimentally observed events and
the $C$ is some constant which is irrelevant 
for our purpose of parameter estimation 
and the determination of confidence regions.  
Here, $n(E)$ is the expected positron energy 
spectrum at Kamiokande or IMB detector
which is computed taking into account
the detector efficiency as well as energy 
resolution in the same way as in Ref. \cite {JNR96}. 
For a combined analysis of the Kamiokande 
and IMB detectors, the likelihood function
is defined as the product of the 
likelihood function for each detector.

We draw in Fig. 2 equal likelihood contours 
as a function of the heavy to light temperature 
ratio $\tau$ on the space spanned by 
$\bar{\nu}_e$ temperature and total neutrino 
luminosity by giving the neutrino events 
from SN1987A observed by Kamiokande and 
IMB detectors \cite {SN1987A}.
In addition to it we introduce an extra 
parameter $\eta$ defined by 
$L_{\nu_x} = L_{\bar{\nu}_x} 
= \eta L_{\nu_e} = \eta L_{\bar{\nu}_e}$
which describe the departure from equipartition 
of energies to three neutrino species and examine the 
sensitivity of our conclusion against the change in the 
SN neutrino spectrum. 
For simplicity, as in Ref. \cite {JNR96}, 
we set the ``effective'' chemical potential 
equal to zero in the neutrino distribution 
functions because we believe that our results 
would not depend much even if we introduce
some non-zero chemical potential. 

At $\tau = 1$, that is at equal $\bar{\nu}_e$ 
and $\nu_e$ temperatures, the 95 $\%$ likelihood 
contour marginally overlaps with the theoretical 
expectation \cite{Janka} represented by the 
shadowed box in Fig. 2.
When the temperature ratio $\tau$ is varied 
from unity to 2 the likelihood contour moves 
to the left, indicating less and less consistency, 
as $\tau$ increases, between the standard 
theoretical expectation and the observed feature 
of the neutrino events after the MSW effect in 
SN is taken into account.
This is simply because the observed energy 
spectrum of $\bar{\nu}_e$ must be interpreted 
as that of the original one of $\bar{\nu}_{heavy}$,
in the presence of the MSW effect in 
the anti-neutrino chanel, which implies that 
the original $\bar{\nu}_e$ temperature
must be lower by a factor $\tau$ than 
the observed one, leading to stronger 
inconsistency at larger $\tau$.

The solid lines in Fig. 2 are for the case 
of equipartition of energy into three flavors, 
$\eta = 1$, whereas the dotted and the dashed 
lines are for $\eta = 0.7$ and 1.3, respectively.
We observe that our result is very 
insensitive against the change in $\eta$.

We conclude that if the temperature ratio 
$\tau$ is in the range 1.4-2.0 as the SN 
simulations indicate, the inverted hierarchy 
of neutrino masses is disfavored by the neutrino 
data of SN1987A unless the H resonance 
is nonadiabatic.\footnote{This result has been 
announced at two international conferences, 
Dark2000 and NOW2000 \cite {DarkNOW}.
} 

For completeness, we summarize the features 
of neutrino events that we expect in 
the three-flavor mixing scheme of neutrinos. 
They differ depending upon the normal or the 
inverted mass hierarchies and on the nature 
of the H and the L level crossings. 

\noindent
{\bf The case of normal mass hierarchy}

There is no level crossing in the antineutrino 
channel, apart from the possibility that 
the solar mixing angle is in the ``dark side'' 
in the parameter space \cite {darkside}, 
$\theta_{12} > \pi/4$, which we do not consider 
in this paper.
Therefore, as far as $\theta_{13}$ is not large, 
which is indicated by CHOOZ result\cite {CHOOZ},   
third state $\bar{\nu}_3$ is essentially decoupled
from the other antineutrinos and 
as far as $\bar{\nu}_e$ signal is concerned, 
the problem can be approximately reduced 
to the two flavor mixing scheme which 
is well explored by the previous works 
\cite {SSB94,JNR96}. 

The conclusion reached can be summarized as follows: 
If the small mixing angle (SMA) MSW is the solution to the solar 
neutrino problem there is only a minor effect because neither 
the vacuum oscillation nor the earth matter effects are effective 
because of small $\theta_{12}$. 
If the large mixing angle (LMA) or 
low $\Delta m^2$ (LOW) MSW or 
vacuum oscillation (VO) is the solution, 
we have a potential trouble because a good 
fraction of $\bar{\nu}_e$ is transformed 
into $\bar{\nu}$$_{heavy}$ and vice versa. 
One can repeat the similar analysis as we 
did for Fig. 2 and would conclude that 
all of these solutions with large mixing angle 
are disfavored \cite {SSB94}, though less 
convincingly than the case of inverted 
mass hierarchy. 

Fortunately, the earth matter effect helps 
us to cure the trouble at least to some extent
for the case of LMA MSW 
solution \cite {SSB94,JNR96}. 
We present in Fig. 3 the result of the 
analysis using the same likelihood method. 
We employ a particular set of parameters of 
the LMA MSW solution and compare the behavior 
of the likelihood contours with and without 
earth matter effect. 
For simplicity, we set $\theta_{13}$ = 0 but our result does not 
change much as long as the parameter is under the CHOOZ bound 
\cite {CHOOZ}.  
In the present analysis we only 
deal with the case of equipartition of energy, $\eta = 1$, because 
we already know that the results are not sensitive to $\eta$.
As one observes in Fig. 3, the earth matter effect cures 
the discrepancy between the likelihood contours and the 
theoretical expectation to some extent, but not completely. 

For the case of LOW MSW as well as VO solutions, 
there is no significant earth matter effect and the results are 
essentially the same as was presented in Fig.5 in Ref. \cite{JNR96}, 
and hence we show no plot for these solutions.

There exist interesting effects in the neutrino 
channel because they have level crossings. We have to discuss 
three cases separately; the H resonance is (a) adiabatic,
(b)moderately nonadiabatic, and (c) nonadiabatic.  

In the case (a) the net effect is 
$\nu_e$-$\nu_{heavy}$ exchange 
independent of the nature of 
the L resonance, as we have discussed 
extensively. The characteristic signature of harder spectrum 
of $\nu_e$ which comes from $\nu_{heavy}$ in neutrinosphere is:

\noindent
(i) Enhancement of forward peaking elastic scattering 
events at high energies which should be observable 
in water Cherenkov detectors \cite {MN90}, and 

\noindent
(ii) Enhanced oxygen-induced events due to 
a steep rise of the cross section at energies 
higher than $\gsim$ 30 MeV \cite {Haxton}, 
which could be separated from the dominant isotropic 
$\bar{\nu_e}$ absorption events due to a moderate 
backward peaking of the events \cite {mina,QF94}. 

In the case (c) one can disregard the third state 
and the problem is the pure two-flavor 
$\nu_e$-$\nu_{heavy}$ transformation. 
Then, if the LMA or LOW MSW is 
the solution to the solar neutrino problem, 
there is a significant conversion 
with similar  experimental 
signatures as the case (a).
We expect that nothing happens in SN in the case 
of VO solution, and about equal mixture 
of $\nu_e$ and $\nu_{heavy}$ 
(for $\sin^2 2\theta_{12} \simeq 1)$
would result due to vacuum oscillations
which would also lead similar but somewhat 
weaker experimental signatures as the case (a).
In the case of SMA MSW solution the conclusion 
depends upon the profile of the outer envelope 
of progenitor star and it is difficult to 
draw definitive conclusions.

In the case (b) in which the conversion 
probability at the H resonance is, say, $\sim 0.5$, 
then one can expect some similar but weaker effects 
with the case (a) independent of the solar neutrino solution, 
and in addition the effects which depend upon the solar 
solutions. 

\noindent
{\bf The case of inverted mass hierarchy}

The structure of the level crossing is very 
simple in this case as shown in Fig. 1b. 
There is the H resonance in the antineutrino 
channel and the L resonance is in the neutrino channel,
since we do not consider the solar parameter 
which is in the ``dark side''. 
Then, the predictions to the neutrinos are 
exactly the same as the ones for case (c) 
in the normal mass hierarchy. 

In the antineutrino channel there is the H resonance 
and we have discussed extensively what happens 
if it is adiabatic, the case (a). 
So we only need to discuss the case (b) and (c).
If the H resonance is nonadiabatic, the case (c), 
the effect of neutrino mixing is the same 
as in the case of normal mass hierarchy. 
In the case of (b), moderately nonadiabatic case, 
$\bar{\nu}_e$-${\bar{\nu}}_{heavy}$ 
transformation occurs 
with the probability $1 - P_H$, and it would imply 
the similar but milder effect than that we have 
obtained with adiabaticity of the H resonance.
If the next galactic supernova is detected 
by Superkamiokande, then we will be able 
to discriminate the moderately nonadiabatic 
case from the adiabatic one. 

After completion of our work we became aware of 
the paper by Lunardini and Smirnov \cite {LS00} 
in which some points related with our work are 
mentioned, but with particular emphasis 
primarily on the detector-dependent earth 
matter effect.

\acknowledgments 

We thank Chung Wook Kim and Korea Institute for 
Advanced Study for the hospitality during our 
visit where this paper was completed. 
We also thank A. Yu Smirnov and C. Lunardini 
for useful discussions. 
This work was supported by the Brazilian funding 
agency Funda\c{c}\~ao de Amparo \`a Pesquisa do 
Estado de S\~ao Paulo (FAPESP), and by 
the Grant-in-Aid for Scientific Research in 
Priority Areas No. 11127213, Japanese Ministry 
of Education, Science, Sports and Culture. 


\begin{figure}[ht]
\vglue 3.0cm 
\hglue -1.0cm 
\centerline{\protect\hbox{
\psfig{file=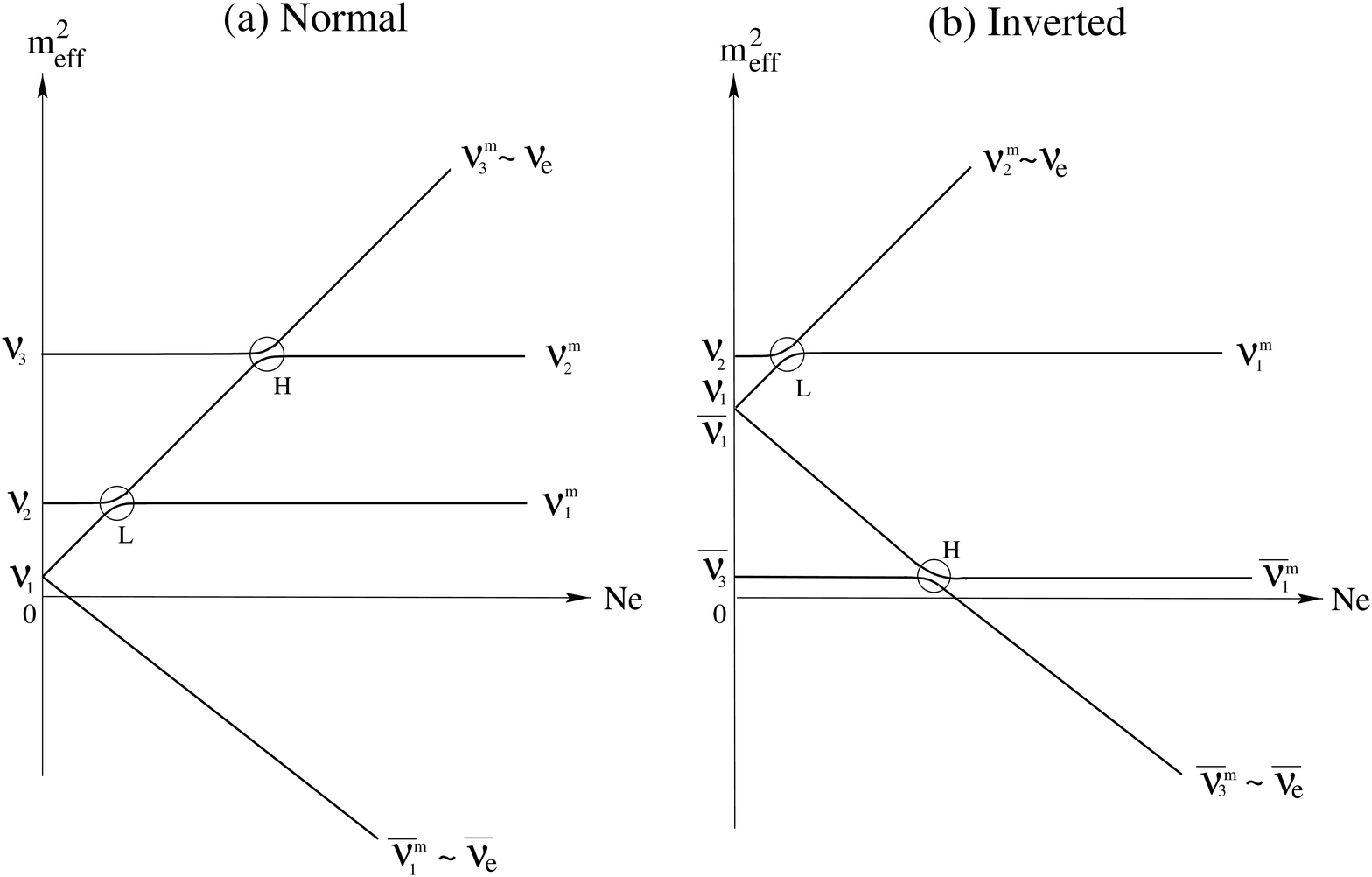,height=11cm,width=17.cm}
}}
\vglue 1.5cm 
\caption{
The schematic level crossing diagram for
the case of (a) normal and (b) inverted mass hierarchies  
considered in this work. 
The circles with the symbol H and L correspond 
to resonance which occur at higher and lower 
density, respectively. 
}
\label{Fig1}
\end{figure}

\newpage

\vglue 1.5cm 
\begin{figure}[ht]
\hglue -1.0cm 
\centerline{\protect\hbox{
\psfig{file=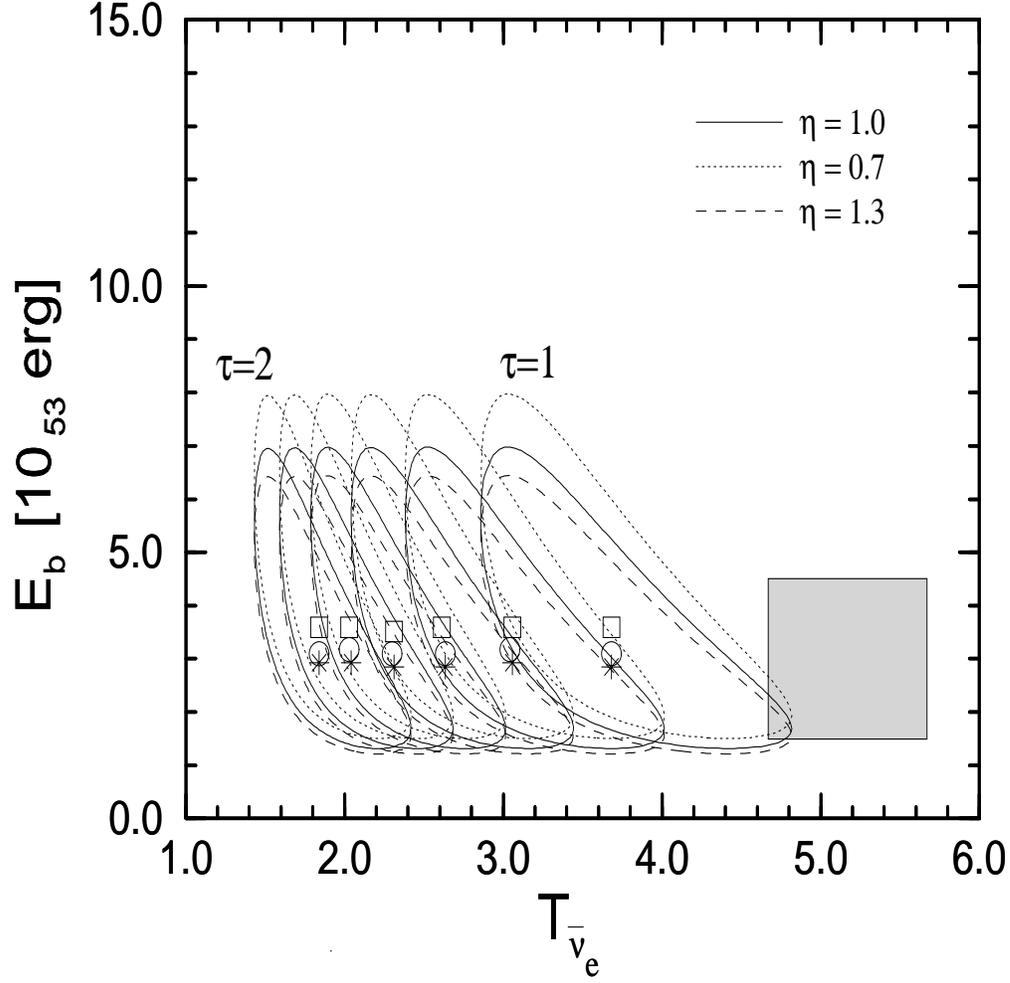,height=14cm,width=14.cm}
}}
\vglue 0.3cm 
\caption{
Contours of constant likelihood which correspond to
95.4 \% confidence regionsfor the inverted mass hierarchy 
under the assumption of adiabatic H resonance.
{}From left to right,
$\tau \equiv T_{\bar{\nu}_x}/T_{\bar{\nu}_e} =
T_{{\nu_x}}/T_{\bar{\nu}_e} =  2, 1.8, 1.6, 1.4, 1.2$ and 1.0
where $x = \mu, \tau$.
Best-fit points for
$T_{\bar{\nu}_e}$ and $E_b$ are also shown
by the open circles.
The parameter $\eta$ parametrizes the departure from the 
equipartition of energy,  
$ L_{\nu_x} = L_{\bar{\nu}_x}= \eta L_{\nu_e} = \eta L_{\bar{\nu}_e}
\ (x =\mu, \tau)$,
and 
the dotted lines (with best fit indicated by open squares) and
the dashed lines (with best fit indicated by stars) 
are for the cases $\eta = 0.7$ and 1.3, respectively.
Theoretical predictions from supernova models 
are indicated by the shadowed box. 
}
\label{Fig2}
\end{figure}

\newpage

\vglue 3.5cm 
\begin{figure}[ht]
\hglue -5.0cm 
\centerline{\protect\hbox{
\psfig{file=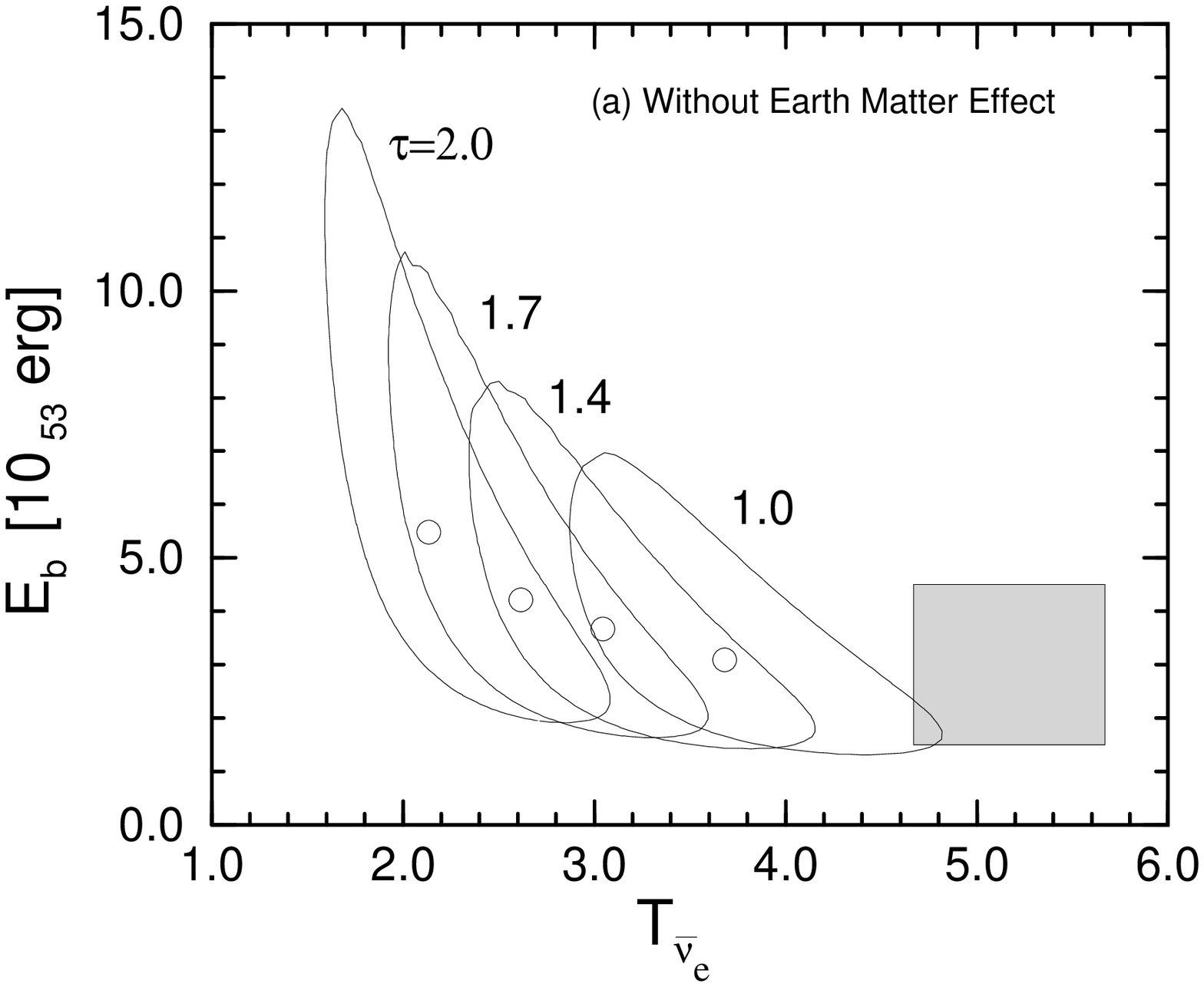,height=8.8cm,width=8.2cm}
}}
\vglue -8.8cm 
\hglue 3.0cm 
\centerline{\protect\hbox{
\psfig{file=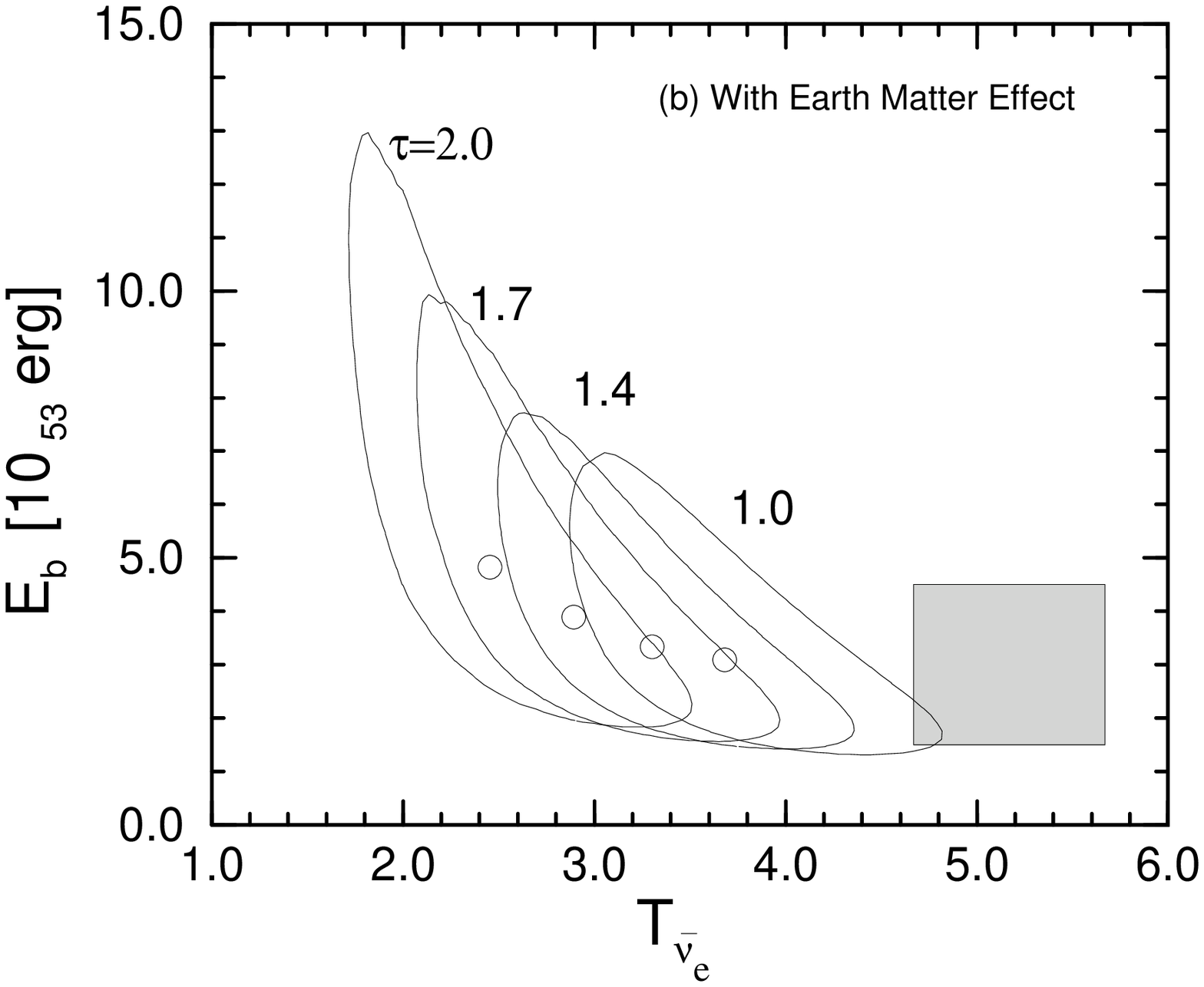,height=8.8cm,width=8.2cm}
}}
\vglue 0.3cm 
\caption{
Contours of constant likelihood corresponding to 
95.4 \% C.L. for the large mixing
angle MSW solution (a) without and (b) with 
earth matter effect. We have taken 
mixing parameters as 
$\Delta m^2 = 3\times 10^{-5}$ eV$^2$ and
$\sin^2 2\theta = 0.8$ for LMA MSW solution. 
}
\label{Fig3}
\end{figure}

\end{document}